\documentclass[amsmath,amsfonts,amssymb,superscriptaddress,twocolumn]{revtex4-2}

\date{\today}
\usepackage{graphicx}
\usepackage{verbatim}
\usepackage{hyperref}
\usepackage{xcolor}
\usepackage{subfigure}% for "subfigure" environment
\usepackage{siunitx}

\begin{document}
\title{Proton-Driven Plasma Wakefield Acceleration\\ for Future HEP Colliders}

\begin{abstract}
\noindent 
We discuss the main elements of a collider facility based on proton-driven plasma wakefield acceleration. We show that very competitive luminosities could be reached for high energy $e^+e^-$ colliders. A first set of parameters was developed for a Higgs Factory indicating that such a scheme is indeed potentially feasible.  There are clearly many challenges to the development of this scheme, including novel RF acceleration modules and high precision and strong magnets for the proton driver.  Challenges in the plasma acceleration stage include
 the ability to accelerate positrons while maintaining necessary emittance and the energy transfer efficiency from the driver to the witness. Since
many exciting applications would become available from our approach, its development should be pursued.

\end{abstract}

\thispagestyle{empty}
\makeatletter
\onecolumngrid
\vspace*{1cm}
\begin{center}
    {\LARGE{\bfseries\sffamily \@title}\par} 
    \vspace*{1cm}
    \includegraphics[width=5cm]{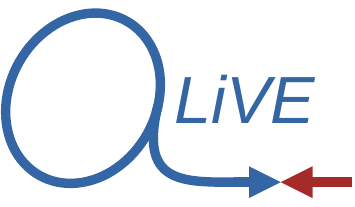}\par
    \vspace*{1cm}
    {
    \large {\bfseries\sffamily Contact Persons}\par
    \vspace*{0.2cm}
    Allen Caldwell, Max Planck Institute for Physics\\\href{mailto:caldwell@mpp.mpg.de}{caldwell@mpp.mpg.de}\par
    \vspace*{0.2cm}
    John Farmer, Max Planck Institute for Physics\\\href{mailto:j.farmer@cern.ch}{j.farmer@cern.ch}\par
              \vspace*{0.2cm}
    Nelson Lopes, GoLP/Instituto de Plasmas e Fusao Nuclear\\ Instituto Superior Tecnico, Universidade de Lisboa \\
    {nelson.lopes@tecnico.ulisboa.pt}\par
            \vspace*{0.2cm}
    Alexander Pukhov, Heinrich-Heine-Universit\"at \\
    {pukhov@tp1.uni-duesseldorf.de}\par
      \vspace*{0.2cm}
    Ferdinand Willeke, Brookhaven National Laboratory \\
    {willeke@bnl.gov}\par
        \vspace*{0.2cm}
    Thomas Wilson, Heinrich-Heine-Universit\"at \\
    {thomas.wilson@hhu.de}\par
    }
    \vspace*{1.5cm}
    \large{\bfseries\sffamily Abstract}
\end{center}
\begin{quotation}
% \noindent\@abstract
\frontmatter@abstract@produce
\end{quotation}
\vfill

\makeatother
\newpage

\setcounter{page}{0}

\title{Proton-Driven Plasma Wakefield Acceleration for Future HEP Colliders}
\author{This document is endorsed by: Carolina Amoedo}
\affiliation{Instituto Superior Tecnico, Lisboa, Portugal}

\author{Osnur Apsimon}
\affiliation{University of Manchester, Manchester, UK}
\affiliation{The Cockcroft Institute, Warrington, UK}

\author{Allen Caldwell}
\affiliation{Max-Planck-Institut f\"ur Physik, Garching, Germany}

\author{Vera Cilento}
\affiliation{CERN, Meyrin, Switzerland}

\author{Moses Chung}
\affiliation{POSTECH, Pohang, Korea}

\author{Arthur Clairembaud}
\affiliation{Max-Planck-Institut f\"ur Physik, Garching, Germany}

\author{G\'abor Demeter}
\affiliation{Wigner Research Centre for Physics, Budapest, Hungary}

\author{Steffen Doebert}
\affiliation{CERN, Meyrin, Switzerland}

\author{John Farmer}
\affiliation{Max-Planck-Institut f\"ur Physik, Garching, Germany}

\author{Brian Foster}
\affiliation{DESY, Hamburg, Germany}
\affiliation{John Adams Institute for Accelerator Science at University of Oxford, Oxford, UK}

\author{Rogelio Tomas Garcia}
\affiliation{CERN, Meyrin, Switzerland}

\author{Edda Gschwendtner}
\affiliation{CERN, Meyrin, Switzerland}

\author{Nicole Hartman}
\affiliation{Technical Univerity Munich, Germany}

\author{Pavel Karataev}
\affiliation{John Adams Institute at Royal Holloway, University of London, UK}

\author{Nelson Lopes}
\affiliation{Instituto Superior T\'ecnico, Lisboa, Portugal}

\author{Alexander Pukhov}
\affiliation{Heinrich-Heine-Universit\"at, D\"usseldorf, Germany}

\author{Jorge Viera}
\affiliation{Instituto Superior T\'ecnico, Lisboa, Portugal}

\author{Carsten P Welsch}
\affiliation{University of Liverpool, UK}

\author{Ferdinand Willeke}
\affiliation{Brookhaven National Laboratory, USA}

\author{Matthew Wing}
\affiliation{University College London, UK}
\affiliation{DESY, Hamburg, Germany}

\author{Thomas Wilson}
\affiliation{Heinrich-Heine-Universit\"at, D\"usseldorf, Germany}

\author{Guoxing Xia}
\affiliation{University of Manchester, Manchester, UK}
\affiliation{The Cockcroft Institute, Warrington, UK}

\author{Giovanni Zevi Della Porta}
\affiliation{CERN, Meyrin, Switzerland}

\begin{abstract}\end{abstract}
% \maketitle

\onecolumngrid
\makeatletter
\begin{center}
    {\Large \bfseries\sffamily \@title\par} 
    \vspace*{0.2cm}
    {\large The ALiVE Collaboration\par}
  % \@author@finish
  %   \groupauthors@sw{%
  %  \frontmatter@author@produce@group
  % }{%
  %  \frontmatter@author@produce@script
  % }
  % \titleblock@produce
  %\maketitle
\end{center}
\makeatother
\vspace*{0.5cm}
\twocolumngrid

%\tableofcontents
%\input{body}
\section{Introduction}

Acceleration in plasma generates interest due to the very strong accelerating gradients it allows. 
 It has been demonstrated that plasmas can be used for accelerating particles to relativistic energies;
gradients well above $1$~GV/m have been achieved.
The advantage of a proton driver is that the energy of the driver is sufficient to reach witness particle energies required for a Higgs Factory and beyond without staging, greatly simplifying the accelerator complex.  In this submission to the EPPSU process, we discuss recent exciting progress using proton-driven plasma wakefield acceleration (PDPWA).

A limitation of PDPWA has so far been the repetition rate of the driver, which limits the achievable luminosity. We discuss approaches to overcome this difficulty, and find that a promising option for a proton driver providing a sufficiently high rate of proton bunches is indeed available.  

A general concern of plasma wakefield acceleration is the control of the bunch emittance during acceleration.  One possibility to resolve 
%avoid 
this is to use conventional accelerators for the positron acceleration, as suggested and pursued by the HALHF collaboration~\cite{ref:HALHF}. A further concern, perhaps most pronounced for PDPWA, is the energy efficiency of the scheme. These and other issues are discussed in a variety of other documents submitted to the EPPSU process (10~TeV design study~\cite{EPPSU2026_tenTeV}, ALEGRO~\cite{EPPSU2026_ALEGRO}, HALHF~\cite{EPPSU2026_HALHF}, LCVision~\cite{EPPSU2026_LCVision}, AWAKE~\cite{EPPSU2026_AWAKE}

We begin our report on the progress of a PDPWA scheme for high energy colliders with an overview of a possible accelerator complex, and then discuss the novel ideas for an effective proton driver.  We then discuss the plasma acceleration stage, including our developing concepts for realizing the necessary plasma.  We finish with an outlook on possible applications.

\section{Accelerator Complex}

\begin{figure*}[htb] 
    \centering
\includegraphics[width=0.75\textwidth]{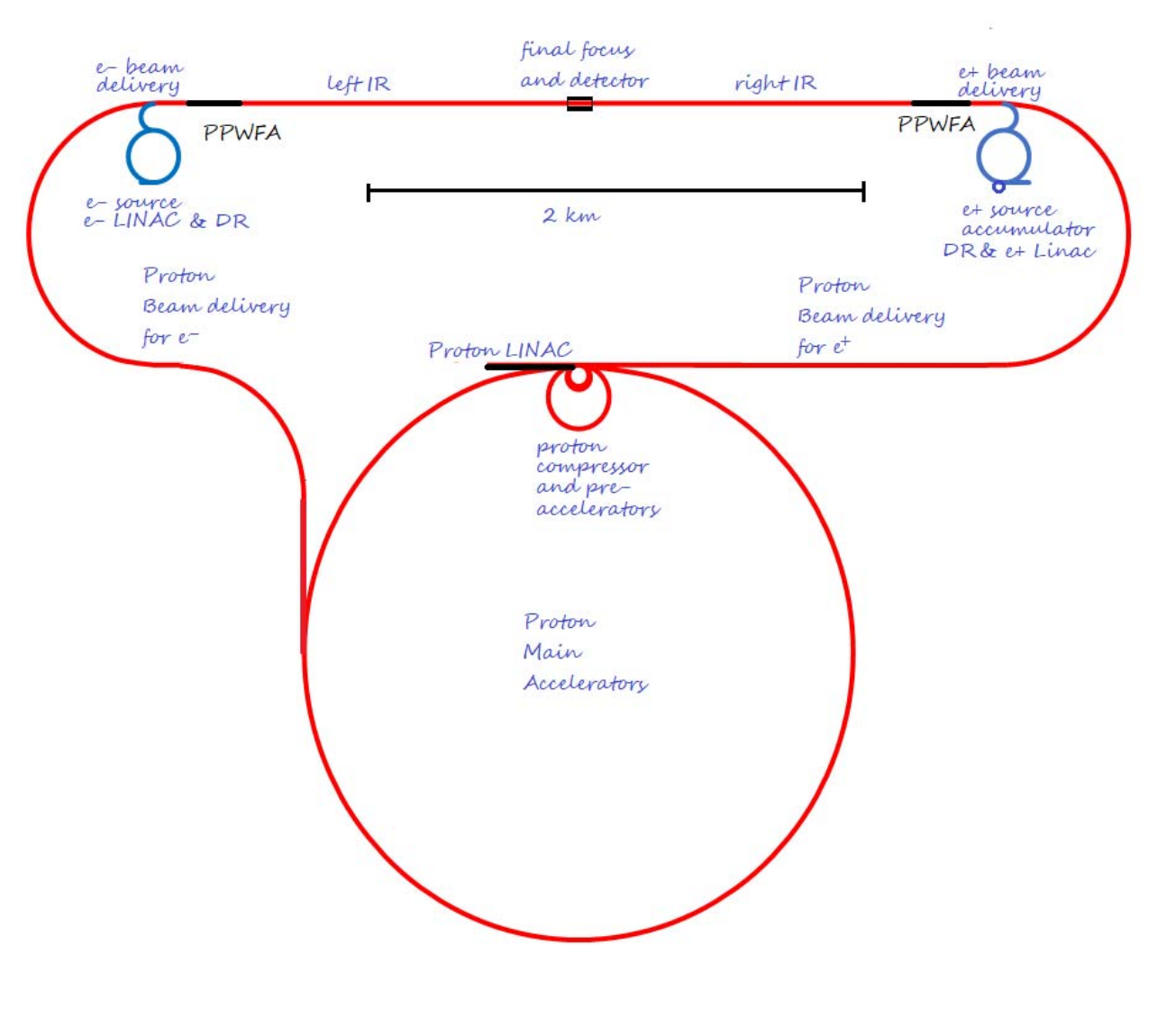}
\caption{\it  Footprint of a PDPWFA Higgs factory facility driven by an FFAG proton accelerator.}
\label{fig:footprint} 
\end{figure*}

An accelerator complex, here for a Fixed Field Alternating Gradient (FFAG) scheme (discussed in the next section) is shown in Fig.~\ref{fig:footprint}.  A continuous sequence of proton bunches is brought to the necessary energy in a ring and then extracted one bunch at a time.  The bunches are then injected into the plasma acceleration sections, trailed by the electron and positron bunches to be accelerated.  The accelerated lepton bunches are then extracted and brought to the interaction point.

%========================================================================================
\begin{table}
\begin{center}
\begin{tabular}{lccc}
\hline
Parameter &  & Value & unit\\
\hline
Beam Energy & $E$ & 125 & \si{\giga\electronvolt}\\
Number of particles per bunch & N & \num{2e10} &\\
$e^-$ bunch length & $\sigma_{e^-,z}$ & 105 & \si{\micro\metre} \\
$e^+$ bunch length & $\sigma_{e^+,z}$ & 75 & \si{\micro\metre} \\
Horizontal $\beta$-function at IP & $\beta_x^*$ & 13 & \si{\milli\metre} \\
Vertical $\beta$-function at IP & $\beta_y^*$ & 0.41 & \si{\milli\metre}\\
Norm. horizontal $e^-$ emittance & $\gamma\varepsilon_{e^-,x}$ & 100 & \si{\nano\metre} \\
Norm. vertical $e^-$ emittance & $\gamma\varepsilon_{e^-,y}$ & 100 & \si{\nano\metre} \\
Norm. horizontal $e^+$ emittance & $\gamma\varepsilon_{e^+,x}$ & 400 & \si{\nano\metre} \\
Norm. vertical $e^+$ emittance & $\gamma\varepsilon_{e^+,y}$ & 400 & \si{\nano\metre} \\
Bunch frequency & $f$ & 7.2 & \si{\kilo\hertz} \\
\hline
Centre-of-mass energy & $E_{cm}$ & 250 & \si{\giga\electronvolt}\\
%Vertical Disruption & D & 73.6 &\\ 
%enhancement factor & $\mathcal{H}$ & 2 &\\
Geometric luminosity      & $\mathcal{L}_\mathrm{geom}$ &  1 & $10^{34}$~\si{\per\square\centi\metre\per\second} \\
Simulated luminosity      & $\mathcal{L}$               & 9.3 & $10^{34}$~\si{\per\square\centi\metre\per\second} \\
Luminosity in the top 1\% & $\mathcal{L}_\mathrm{1\%}$ & 0.34 & $10^{34}$~\si{\per\square\centi\metre\per\second} \\
\hline
\end{tabular}
\end{center}
\caption{Luminosity parameters for a Higgs factory based on proton-driven plasma wakefield acceleration.}
\label{tab:Table-1}
\end{table}

The luminosity  of the $e^+e^-$ collider is
\begin{equation*}
\mathcal{L} = \frac{f_b \cdot N_{e+}\cdot N_{e-} \cdot\mathcal{H}}
{2\pi \sqrt{(\varepsilon _x^-+\varepsilon _x^+) \cdot \beta_x^{*} \cdot (\varepsilon_y^-+\varepsilon_y^+) \cdot \beta_y^{*}}}
\end{equation*}
where $N_{e+,e-}$ are the number of leptons per bunch, $f_b$ is the bunch frequency, $\varepsilon_{x,y}$ are the lepton emittances, and $\beta_{x,y}^*$
are the $\beta$-functions at the collision point. $\mathcal{H}$ describes disruption, pinch and hour glass effects. 
In the next section, we will use the Higgs Factory as an example application of a PDPWA-based facility. The parameters for one version of such an $e^+e^-$ Higgs Factory are summarized in Table \ref{tab:Table-1}.  

In the following, we discuss the main elements needed in the realization of such a facility.  These are the conventional (non-plasma based) accelerators needed to provide the driver and witness beams, the plasma acceleration, and the development of suitable plasma cells. The design of a beam delivery system to the IP is left for future work. We begin with the conventional accelerator elements.

\section{RF-based Accelerators}

We first outline the elements of the accelerator complex based on conventional RF acceleration. The facility has a number of constituents as shown in Fig.~\ref{fig:footprint}.

\begin{itemize}
\item the electron injector consists of a 1~\si{\giga\electronvolt}, 1.2~\si{\giga\hertz} superconducting LINAC, and a 762~\si{\metre} damping ring
\item the positron injector consisting of positron pre-LINAC, positron production target, positron accumulator, 1~\si{\giga\electronvolt} positron LINAC and  positron damping ring
\item the proton injector consisting of the source, the 10~\si{\giga\electronvolt} LINAC and two compressor rings
\item the proton accelerators consisting of  three or four circular accelerator rings
(depending on the accelerator scheme)
\item the proton delivery beam lines with a compressor chicane
\item the plasma wakefield accelerating channels
\item the interaction region with colliding beam detector
\end{itemize}
We summarize the injector concepts without giving design details.

\subsection{Lepton Bunch Preparation}
The lepton parameters are similar to the ones of linear colliders (see for example  \cite{ref:CLIC_CDR}) and we are leaning strongly on the design of particle sources, positron production 
and damping rings that have been carried out for linear colliders. 

The generation of polarized electron (and positron) bunches of $N_e=$\num{2e10} particles and a normalized emittance of $\varepsilon_N = 3.4$~\si{\micro\metre} with a repetition rate of 7.2~\si{\kilo\hertz} is well within the state of the art of polarized electron sources \cite{ref:Wang22}. The beam is accelerated by a 1~\si{\giga\electronvolt} superconducting LINAC followed by a 762~\si{\metre} damping ring (DR). The requirement of  a normalized electron emittance of $\varepsilon_N= 100$~\si{\pi\,\nano\metre\,\radian}  corresponds to a DR equilibrium emittance of $< 50$~\si{\pi\,\pico\metre\,\radian} which is small but its feasibility is supported by the design of several small light sources under construction or under study (see reference\cite{ref:Shin} for an overview). The DR should have  a transverse damping time of  $5$~\si{\milli\second}, which requires damping wigglers ($eU_{loss}/turn=0.5$~\si{\mega\electronvolt}). Each of the $N_{be} = 127 $ damping ring bunches circulates for 3.4 damping times before being extracted and transferred to the plasma wakefield acceleration channel.  Several schemes have been proposed for polarized positron production \cite{ref:Rinolfi} the details of which we are not going to discuss at this point. Positrons need to be accumulated in a small accumulator ring. No accumulator is required for electrons in this scheme because of the uniform bunch frequency of 7.2~\si{\kilo\Hz}.

\subsection{Proton Bunch Preparation and Acceleration}\label{sec:protons}
%This will be discussed in the following %sections. 
%
%\subsection{Proton Acceleration for high %bunch Rate}
For a competitive high luminosity ($L>10^{34}$~\si{\per\square\centi\metre\per\second}) $e^+e^-$ Higgs factory, the proton drive bunches that generate the plasma wakefield for lepton acceleration should ideally have a beam energy of about 500~\si{\giga\electronvolt}, a high beam intensity, $N_p \simeq \num{1e11}$ protons per bunch, short bunch length, $\sigma_b \ll 1$~\si{\milli\metre}, and a high repetition rate $f_b$. The repetition rate should only be limited by the power $P_b$ needed to accelerate the protons. The envisioned bunch frequency of $7.2$~\si{\kilo\hertz} for each of the two beams corresponds to a beam power of  $P_b\simeq 62.5$~\si{\mega\watt} per beam which we consider as an administrative limit. 
The required proton rate of $2\times \num{0.72e15}$/\si{\second} is well within the capacity of present day proton sources (100~\si{\milli\ampere} at 5\% duty factor corresponding  to \num{3e16}/\si{\second}~\cite{ref:SNS-Operations}). It is, however  challenging to deliver these protons in bunches of $10^{11}$ particles at $E=500$~\si{\giga\electronvolt} at a steady rate of $f_b = 14.4$~\si{\kilo\hertz}. 

It is optimal to deliver proton bunches equally spaced at a constant rate. 
We only consider  multi-turn, circular accelerators.

\subsubsection{Rapid Cycling Synchrotrons}
We first  consider fast cycling synchrotrons. To obtain equal proton bunch spacing we assume that the entire beam is transferred at top energy in one turn to a stretcher ring from which proton bunches are extracted at a rate of 14.4~\si{\kilo\hertz} while the synchrotron is cycling.  

A high repetition rate  is enabled by rapidly changing magnetic guide fields $B$ and short injection time. Many factors limit the product $B\cdot \dot{B}$. In superconducting magnets some are of fundamental nature and are hard to mitigate (see for example \cite{ref:Kirby2008}). The SIS100 super ferric magnet system constitutes the present state of the art of mass-produced fast ramping superconducting magnets \cite{ref:SIS100}. The SIS100 dipole has a peak field of 2$T$ and the magnets are ramped in about 0.4~\si{\second} corresponding to $B\cdot\dot{B}$= 5~\si{\square\tesla\per\second}. Similar values $B\cdot\dot{B}$ were obtained when testing the SIS300 superconducting magnet prototype magnet \cite{ref:SIS300}.
%that has a field of 4.5 $T$ and can be ramped %with $\dot{B}=1T/s$   which  corresponds to %$B\cdot{B}$= 4.5 $T^2/s$. 

Recent R\&D work on high temperature superconductors (HTS) is quite promising. At FERMILAB, an HTS  prototype magnet made with  REBCO high temperature conductor was operated with a temperature of 6~\si{\kelvin} at magnetic fields between $B = 0\,$ to $0.4$~\si{\tesla} \cite{Piekarz:2019yis}. The coil design supports the  format of the HTS  and optimum cooling. In the test measurement, the maximum $\dot{B}$ was close to 280 $T/s$ (at $B=0$~\si{\tesla}) and a maximum $B\cdot\dot{B}$ of 50~\si{\square\tesla\per\second} was achieved. While this is very encouraging,  a large R\&D effort is required to arrive at a mass producible magnet that significantly exceeds present limitations of fast ramping superconducting magnets. 

Extrapolating the present state of the art  it appears possible that in 10 years  $B\cdot\dot{B}$ of 20~\si{\square\tesla\per\second} will eventually be achieved in mass-produced (2-3)~\si{\tesla} magnets.  Considering a 500~\si{\giga\electronvolt} proton driver with a circumference of $\approx  6900$~\si{\metre} and a bending radius of 750~\si{\metre},  a peak magnetic field of $B_{peak}$ = 2.2~\si{\tesla}, a ramp cycle frequency of $f_r=1/2\cdot (B\cdot\dot{B})_{max}/B_{peak}^2  = 2$~\si{\hertz} would be achieved.  With $N_b= $ 1000, (700~\si{\milli\ampere} beam current) in the ring, we  project an achievable bunch frequency of  $f_b=2.1$~\si{\kilo\hertz}  falling short of the stated goal of $14.4$~\si{\kilo\hertz}. We also need to take the injection time of the synchrotron into account. Assuming a $1$~\si{\giga\electronvolt} $H^-$-LINAC with a beam current of $100$~\si{\milli\ampere} at 0.3\%
duty factor, with charge exchange injection into a low energy 7~\si{\giga\electronvolt} booster 
%ramping 
with 60~\si{\hertz} cycle rate (comparable to the DESY-I synchrotron \cite{ref:DESY-I:1988ffj}), followed by a 50~\si{\giga\electronvolt} medium energy booster with $900 $~\si{\metre} circumference and a ramp rate of 8~\si{\hertz}
($B\cdot\dot{B}=21$~\si{\square\tesla\per\second}), the injection time in the 500 ~\si{\giga\electronvolt}, 6900~\si{\metre} synchrotron is 
1.5~\si{\second}. Thus, even faster ramping will not 
provide a 
%nearly 
sufficient proton bunch injection rate.

We conclude that even with advances in magnet technology in a foreseeable time, it is unclear whether a rapid cycling synchrotron scheme will provide a proton bunch repetition rate of $f_b =$ 14.4 $kHz$, that is only limited by the power required for acceleration. While solutions to speed a proton production are under study, we are motivated to search for alternatives. 

\subsubsection{Fixed Field Alternating Gradient Accelerator (FFAG) 
%for the Proton Driver
}

FFAGs have been proposed already in the early days of high energy particle accelerators \cite{ref:Simon56:1956ffj} but have not played a large role in their development. In an FFAG,
% like in a cyclotron, 
the magnetic guide and focusing fields are static. Focusing and bending  occurs in  strong, alternating-gradient  combined function magnets.  During acceleration, the beam is moving radially  inside the magnet aperture into regions of larger bending and stronger focusing fields. FFAGs
%(like cyclotrons) 
provide a  continuous stream of high energy bunches. 

We consider a scaling FFAG for the proton driver. The beam optics
%(including tunes)
do not change during acceleration. 
This supports acceleration up to  500~\si{\giga\electronvolt} during many turns ($\approx 700$) which limits the RF voltage to reasonable values. 
Constant optics is achieved by an exponential dependence of the guide field $B_y(x)=B_y(x_0)\cdot \exp{(x/\Delta x)}$ on the radial coordinate $x$. The focusing strength $k = e\cdot c\cdot (\mathrm{d}B_y/\mathrm{d}x)/E= (\mathrm{d}B_y/\mathrm{d}x)/(B_y\rho)= 1/(\rho\cdot \Delta x)$ ($\rho$ is the constant bending radius) 
%Thus
$k $ being controlled by the parameters $\Delta x$ and $\rho$. 
The change in energy is determined by the RF-Voltage and the slowly varying RF phase during acceleration. 

A stable beam optics requires alternating gradients. The guide field of the defocusing magnets needs to have the same dependence on $x$ than the focusing one. Since the gradient of the defocusing magnet has the opposite sign than the focusing one, the guide field of the defocusing magnet has an opposite sign as well. We can bend and focus stronger in the focusing magnet than in the defocusing one to obtain a net radially-inward bending, but optical stability limits the ratio of integrated focusing $kl$ to $R=(kl)_{defoc}/(-kl)_{foc}= 0.65$  (field strengths are kept the same but magnet lengths vary $l_{defoc}=R\cdot l_{foc}$)
%$ = 0.65\cdot l_{foc}$). 
This implies that scaling FFAGs use magnetic field very inefficiently since inward bending magnets need to make up for the outward bending ones. The required peak magnetic field becomes  $B_{peak}= (1+R)/(1-R)\cdot B_{average}= 4.7\cdot B_{average}$ which severely limits the energy reach of the FFAG.
We explore the FFAG concept for generating a $500$~\si{\giga\electronvolt} proton drive beam starting from the following parameters: 
The beam of bunches with $N_p=10^{11}$ protons is accelerated from 150 $\si{\giga\electronvolt}$ to 500\, $\si{\giga\electronvolt}$ in a $6900$~\si{\metre} ring and a bending radius of  $\rho =\pm 163\, m$  which corresponds to a magnetic field range of 3.07 -- 10.22~\si{\tesla}. A fill factor of  $\eta = 0.7$ provides space for RF-cavities and other accelerator components. The RF voltage is $U_{RF}=$\num{7e8}~\si{\volt}, the RF frequency is $f_{RF}=52.906$~\si{\mega\hertz}
%($h=1100$) 
and the RF phase at injection is close to $\phi_{RF}(E_{inj})= -\pi/2$.  The beam is accelerated in about 700 turns. A new bunch is injected every 3rd turn (69.5~\si{micro\second}) and every 3rd turn, a full energy bunch is extracted from the ring. This corresponds to a bunch frequency of 14.4~\si{\kilo\hertz}. There are 230 bunches with energies ranging from 150 ~\si{\giga\electronvolt} to 500~\si{\giga\electronvolt} in the ring at any time. The power transmitted to the beam is 125~\si{\mega\watt} which limits the bunch frequency. The beam current is about $I_p \simeq 160 $~\si{\milli\ampere}.  The lattice is composed of a focusing/defocusing structure of combined function magnets. The magnetic field varies as $B_y(x)= B_{0}\cdot \exp{(x/\Delta x)}$, $\Delta x = 10$~\si{\milli\metre}. The focusing parameters are $k
%=e\cdot c\cdot (dB/dx)/E 
= \pm 0.06$~\si{\per\square\metre}. The lengths of the focusing and defocusing magnets 
%(sector magnets assumed) 
are $l_f=1.951$~\si{\metre} and $l_d= 0.65\cdot L_f= 1.268$~\si{\metre}.  The betatron phase advances per cell are $\Delta\Psi_x= 0.37\cdot 2\pi,\, \Delta\Psi_y= 0.093\cdot 2\pi$ and the $\beta$-function are in the range of  0.85~\si{\metre} $< \beta_x <$ 5.2~\si{\metre} ; 9.7~\si{\metre} $< \beta_y <$ 31.2~\si{\metre}. The nearly constant dispersion is $\simeq$ 10~\si{\milli\metre} and the momentum compaction factor is very small, $\alpha_{mc}=\num{7e-6}$. The lattice is nearly isochronous. 

\begin{figure}
\centering
\includegraphics*[width=0.5\textwidth]{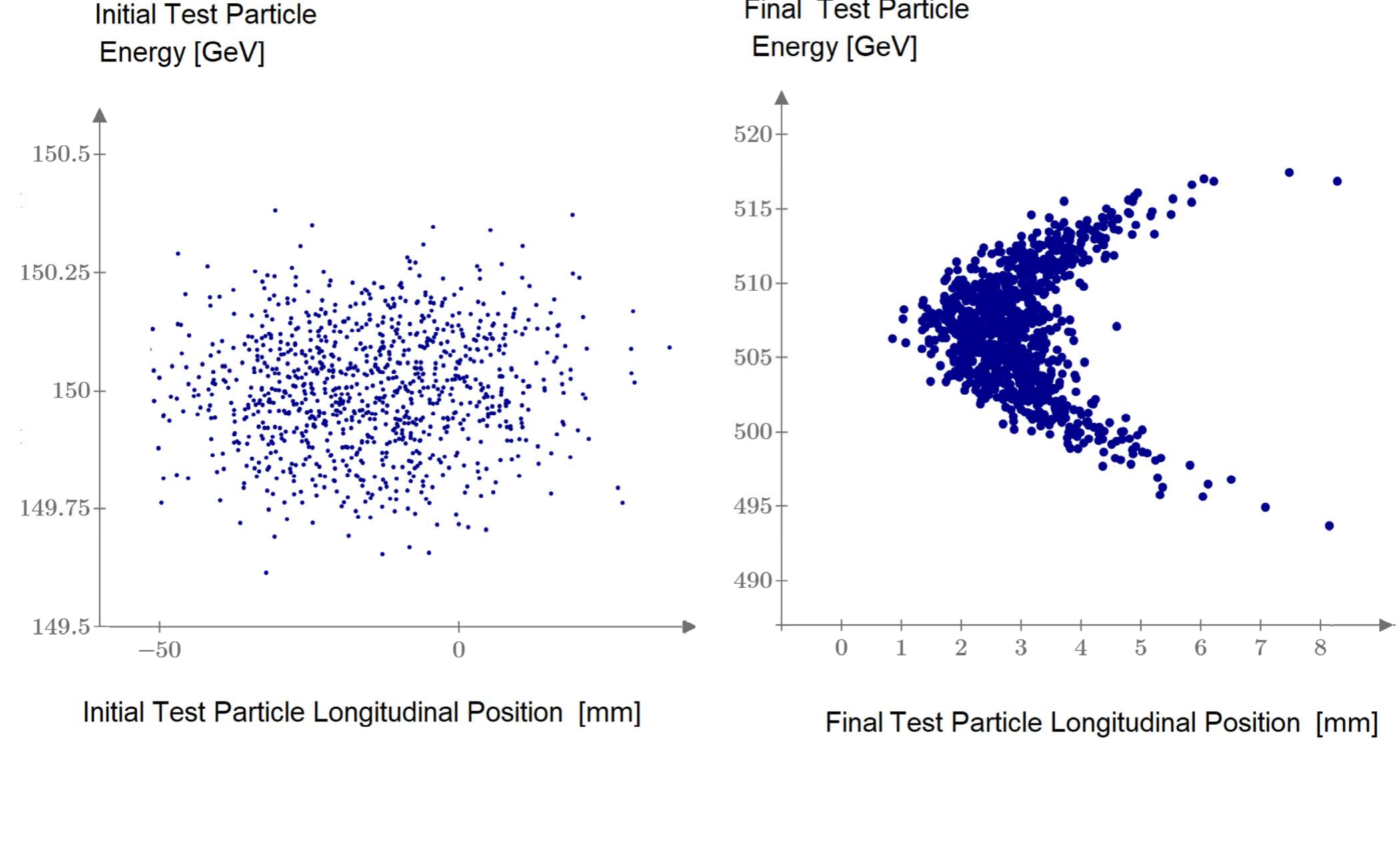}
\caption{Longitudinal phase space of protons represented by 1000 Gaussian-distributed test particles before injection and after extraction and compression. The final rms bunch length is less than 1~\si{\milli\metre}. }
\label{fig:bunch}
\end{figure}

The 
%small 
RF phase varies during acceleration due to changes in the beam velocity and 
%due to 
the outward radial shift towards larger magnetic fields $\Delta s= 2\pi \, \Delta x$. Both effects partially cancel. Thus circumference, energy range, magnet aperture, and phase stability during acceleration are correlated.%(see Figure \ref{Figure 1}) 
%and the RF phase remains in the accelerating regime (see also Figure \ref{Figure 2}). 
 The maximum beam energy of 500~\si{\giga\electronvolt}, limited by the magnetic field, is reached after 688 turns.
 %(see Figure \ref{Figure 3}). 
%
%\begin{figure}
%\centering
%\includegraphics*[width=0.8\textwidth]{pictures/change_in_revolution_time_vs_Beam_Energy.pdf}
%\caption{Revolution versus Beam Energy: blue trace: variation due to path lengthening, red trace: variation due to %velocity change, green: resulting revolution times }
%\label{Figure 1}
%\end{figure}
%
%
%\begin{figure}
%\centering
%\includegraphics*[width=0.8\textwidth]{pictures/evolution of RF-phase.pdf}
%\caption{Change of the RF phase during acceleration staring at $\Psi = 0$, exceeding $\Psi = \pi /2$ only for a short %time and remaining 
%just below  $\Psi = \pi /2$ for most of the acceleration exploiting the EF voltage fairly efficiently that way.}
%\label{Figure 2}
%\end{figure}
%
%\begin{figure}
%\centering
%\includegraphics*[width=0.8\textwidth]{pictures/Beam Energy versus Turns.pdf}
%\caption{Beam Energy of a center particle versus Number of Turns: The maximum beam energy that is limited by %magnetic field  is reached after 688 turns. At this point the beam is extracted. }
%\label{Figure 3}
%\end{figure}
%
The longitudinal dynamics in this lattice is unusual. The  synchrotron frequency is small (25 $Hz$) due to the small momentum compaction factor. Thus the synchrotron oscillation phase advance over the entire acceleration is  $\Delta\Psi_s <  \pi$. The bucket height is very large for the same reason and because of the large RF voltage. We assume aggressive, but demonstrated longitudinal parameters for the protons with a beam area of  $A_{bunch} = \pi\, \Delta E \cdot \Delta t = 0.03$~\si{\electronvolt},  ($\Delta E/E$ = 0.001, $\Delta t = 67$~\si{\pico\second}). The injected beam is mismatched to the RF parameters and is strongly diverging linearly in both phase and energy spread thereby preserving the longitudinal emittance. If we transport this beam after extraction through a chicane with $R_{56}= 0.91$~\si{\milli\metre}, a rms bunch length of $\sigma_b = 0.7$~\si{\milli\metre} is achieved. The beam energy spread is $\simeq 30$~\si{\giga\electronvolt}.
Figure \ref{fig:bunch} shows the longitudinal phase space before injection and after passing through the chicane. %The Figure \ref{Figure 5} depicts resulting longitudinal density distribution.
%
%\begin{figure}
%\centering
%\includegraphics*[width=0.7\textwidth]{pictures/longitudinal-density-distribution.pdf}
%\caption{Longitudinal distribution functions of proton bunches after extraction and compression.}
%\label{Figure 5}
%\end{figure}
%
Two chains of proton bunches are required to 
drive the electron and positron acceleration channels. 
%The proton source has sufficient 
%capacity to produce a second proton bunch chain within the same time. %Only the compressor ring needs to be duplicated. Injection into the %FFAG chain must occur at a $30\,kHz$ rate with a bunch distance of %$33\, \mu s$ which reduces the bunch spacing in the FFAG rings to $50 %\, ns$.
Two fast kicker magnets with a rise time of 100~\si{\nano\second} will extract the bunches into the two proton driver transport lines with a rate of 7.2~\si{\kilo\hertz} respectively. 
%Once a bunch reaches top energy of each stage, it will be extracted by %a fast kicker magnet with $50\, ns$ rise time. 
%Such kickers are quite challenging.

The larger magnet aperture $\Delta x$ in the lower energy rings provides larger change in path length $2\, \pi \,\Delta x$ to balance the larger change in velocity $\Delta v \simeq c\cdot (1/\gamma_{ini}^2-1/\gamma_f^2)$ during acceleration.

The FFAG design requires a considerable effort of optimization to minimize the requirements on accelerator hardware. 
The complete accelerator lattice with transition from the arcs to straight sections remains to be designed.
The accelerator components require a dedicated R\&D program. Magnetic peak fields are quite strong ($B_{peak}= 10.2$~\si{\tesla}). The shape of the field is unusual. It is straightforward to lay out a  superconducting coil configuration consisting of nearly rectangular blocks that provides a nearly exponential field with relative field errors in the order of $10^{-3}$ and gradient errors in the order of $10^{-2}$. The design of the entire magnet system however 
%with magnetic shielding, flux return, conductor return loops, magnet-to-magnet connections, management of forces and stresses, etc  
requires a novel superconducting magnet development program. 
The 52~\si{\mega\hertz} superconducting RF system that provides the accelerating voltage of 700~\si{\mega\volt} is quite unconventional and requires R\&D. 

The final stage of acceleration requires injection of proton bunches at a rate of 14.4~\si{\kilo\hertz} that is to be provided by the injectors. The following scheme is envisioned: 
Following a 10 \si{\giga\electronvolt} proton LINAC operating with 14.4~\si{\kilo\hertz} and two compressor rings there are 4 FFAGs:
\begin{itemize}
\item A $400$~\si{\metre} FFAG from $10$~\si{\giga\electronvolt}-to $15$~\si{\giga\electronvolt} requiring
a maximum field of $B=6.62$~\si{\tesla} and $7$~\si{\centi\metre} horizontal aperture,
\item a $1200$~\si{\metre} FFAG from $15$~\si{\giga\electronvolt}-to $50$~\si{\giga\electronvolt} requiring
a maximum field of $B=5.86$~\si{\tesla} and $18.4$~\si{\centi\metre} horizontal aperture,
\item a $6900$~\si{\metre} FFAG from $50$~\si{\giga\electronvolt}-to $150$~\si{\giga\electronvolt} requiring
a maximum field of $B=3.05$~\si{\tesla} and $10.4$~\si{\centi\metre} horizontal aperture 
\item the final $500$~\si{\giga\electronvolt} acceleration stage discussed above
\end{itemize} 
% The straight section lattice will most likely be constituted of section with 
%equally strong alternating bendings which provide the freedom to vary the focusing for the purpose of optical matching %from the densely populated arcs to the straight sections with empty straight sections for RF and auxiliaries. 

We summarize our findings on FFAGs to accelerate the proton drive beam for plasma wakefield generation:
The FFAG approach appears to be a possible path to provide high intensity protons with a large bunch rates and strongly compressed proton bunches. 
The proton bunch rate is  of $14.4$~\si{\kilo\hertz} in our example is limited by the power that is transferred to the proton beam. 
%and could be as large as 50 $kHz$.  
The required accelerator components  are beyond state of the art and require a  comprehensive R\&D program on novel superconducting magnets and superconducting cavities.

\subsection{Beam Delivery System and Interaction Point}\label{sec:IP}
Colliders typically make use of flat beams, which minimize disruption~\cite{Schulte-flat}.  However, acceleration of flat beams in plasma is challenging, as nonliearities in the plasma response can cause coupling between the transverse planes~\cite{ref:diederichs_flat}.  In this work, we assume round beams accelerated in plasma, but assume that suitable beam manipulations can be carried out to give a somewhat flattened focus.  We therefore take beams with equal emittances in the horizontal and vertical planes, and assume that the $\beta^\ast$ values from the ILC~\cite{ref:ILCpars} can be attained with suitable beam manipulations in the BDS.  The beam parameters remain aspirational, assuming negligible emittance growth for the electron beam, and an emittance growth of a factor 4 for the positron beam.  Other beam parameters in Table~\ref{tab:Table-1} are taken from the simulation studies discussed in Section \ref{sec:HiggsPaper}, with the added assumption that moderate optimization via bunch shaping would allow the population of the witness bunches to be doubled, and an energy spread of 0.1\% to be achieved.

The relatively small aspect ratio of the beams at the IP leads both to significant disruption and enhancement, as the electron and positron beams mutually focus, with the resulting extreme fields leading to significant QED effects.  Simulations of the IP were carried out with GUINEA-PIG~\cite{Schulte-Guinea}, and show that despite the strong disruption, the absolute luminosity in the top 1\% energy range remains competitive.  The extent to which detectors can tolerate such a large background remains an area of future investigation, but we note that developments for the Luminosity upgrade of the Large Hadron Collider should prove beneficial here.  The strong enhancement in this regime appears to make the scheme more robust against the increased positron emittance, although further studies on this are required.

\section{Proton-driven Plasma Acceleration}
% \subsection{AWAKE scheme}
PDPWA was initially considered assuming a short proton bunch~\cite{ref:Caldwell-protondriven} as the driver of plasma wakefields.  As short proton bunches are not available today, an R\&D program was developed~\cite{ref:Path} based on the concept of the self-modulation of a long proton bunch~\cite{ref:Modulation}.  The AWAKE Collaboration has carried out a very successful experimental program based on this approach.  A recent summary of AWAKE results, including prospects for particle physics applications, are discussed in~\cite{ref:AWAKE_future}.  The current status of the AWAKE project and future plans are discussed in a separate submission to the EPPSU process~\cite{EPPSU2026_AWAKE}.

The 2020 update of the European Strategy for Particle Physics~\cite{espp2022} explicitly called for the development of new techniques to increase the luminosity of the AWAKE scheme.  Studies on this theme have been carried out~\cite{pwfa-farmer-multibunch}, and offer a path to increase the luminosity of AWAKE for fixed-target experiments.  %However, self-modulation ultimately limits the efficiency with which energy can be transferred from the driver to the witness.

In this submission to the EPPSU process, we return to the concept of a short proton driver, as it represents a natural next step to further increase the luminosity of proton-driven wakefield schemes, relevant for future collider projects.
%We now discuss the lepton acceleration in a plasma driven by a short, high-energy, proton bunch.
We first review the general concepts and then discuss important aspects that remain to be demonstrated. We then focus on the plasma cell technology and first ideas on realizing the necessary plasmas.

\subsection{Short proton driver}\label{sec:shortdriver}

\begin{figure}
    \centering
    \includegraphics[width=0.8\linewidth]{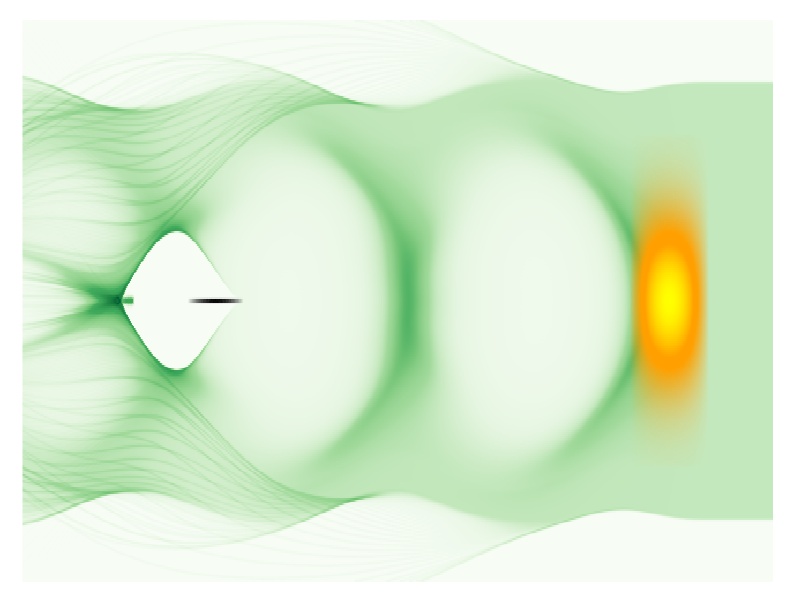}
    \caption{A short proton driver (yellow/orange), propagating to the right, excites a density wave in the plasma electron density (green).  The associated wakefields can be used to accelerate a trailing witness, in this case an electron bunch (black), to high energy.}
    \label{fig:PDPWFA_clipart}
\end{figure}

As a short proton bunch propagates through plasma, it perturbs the plasma electrons, setting them into motion and generating a plasma wave. For a highly relativistic proton drive bunch, the electric field experienced by the plasma electrons is primarily transverse to the proton beam's direction. This causes the electrons to oscillate at a characteristic plasma frequency, given by  

$$
\omega_p = \sqrt{\frac{n_p e^2}{\varepsilon_0 m}},
$$  
where $n_p$ is the plasma electron number density, while $-e$ and $m$ represent the electron charge and mass, respectively. The permittivity of free space is denoted by $\varepsilon_0$. Due to their significantly larger mass, plasma ions move comparatively little. The created wakefield has a large longitudinal field component that can be used to accelerate witness bunches.  

Initially, the oscillating electrons move toward the axis of the proton beam. As they pass through one another, they form a region of reduced plasma electron density, which in turn generates intense electric fields. At a given plasma density, this pattern co-moves with the proton bunch at its velocity. In addition to producing large accelerating gradients on the order of $\sim mc\omega_p/e$, the wakefield structure also provides strong transverse focusing fields.  

A key consideration in proton-driven plasma-based acceleration is the phase slippage between the electron or positron witness bunch and the drive bunch. Since electrons and positrons travel nearly at the speed of light, while protons move slightly below this limit, their relative phase shifts over time. However, this dephasing can be mitigated by carefully controlling the plasma wavelength. By varying the plasma density along the acceleration length~\cite{ref:Katsouleas,ref:PukhovDephasing}, one can compensate for phase slippage, ensuring that the witness bunch remains in the accelerating phase. Consequently, dephasing concerns can be effectively managed through plasma density tailoring.  

\subsection{Accelerating Positrons}\label{sec:positrons}
Plasma wakefield acceleration is highly effective when both the driver and witness particles are negatively charged. In this scenario, a short, high-current driver generates a ``bubble'' or ``blowout'' region~\cite{pwfa-rosenzweig-blowout}, where all plasma electrons are expelled. The resulting positive charge of the remaining ions creates focusing and accelerating fields that efficiently transfer energy to a negatively charged witness.% This bubble regime acts as a highly efficient energy converter from driver to witness.

However, the situation is far less favorable for a positively charged witness, such as a positron bunch. In the strongly nonlinear wake, there exists only a very short spatial region where the fields provide both focusing and acceleration for positrons. This region is highly sensitive to even minor fluctuations in the plasma density profile and driver charge, making positron acceleration in the bubble regime impractical.  

A quasi-linear wakefield with moderate amplitude offers a more symmetric profile, enabling the acceleration of both electrons and positrons. However, the energy transfer efficiency from driver to witness is inherently limited in this case. A low-emittance witness is strongly focused by the wake’s fields, leading to a high local density. Here, the asymmetric nature of plasma as an accelerating medium becomes critical. A negatively charged witness generates its own mini-bubble~\cite{olsen-emittance}, helping to preserve the emittance of an electron bunch. % However, energy efficiency remains constrained, as the witness can only extract wakefield energy within the mini-bubble’s radius. 

In contrast, a positron witness %does not generate a mini-bubble but instead
attracts the surrounding plasma electrons, forming a localized electron density spike. Unlike background ions in the bubble regime, which move comparatively little, this electron spike is much more fragile, posing significant challenges for positron emittance preservation.  Positron acceleration has been identified as one of the ``grand challenges'' of a collider scheme based on plasma-wakefield acceleration~\cite{EPPSU2026_tenTeV}, and several groups within the plasma physics community are engaged in efforts to find more robust schemes positron acceleration, generally making use of structured plasmas~\cite{cao-positronreview}.

%Despite these challenges, our preliminary simulations indicate that positrons can still be accelerated to high energies with acceptable beam quality in the quasi-linear regime. [!!!] 
%Positron beams with few-micron emittance can be accelerated in the quasilinear regime with moderate efficiency~[cite Hue], but new methods are required to accelerate beams with collider-relevant witness quality.  
%  Studies are also underway to better quantify the impact of positron beam quality in high-disruption scenarios. 

\subsection{Energy scaling}\label{sec:energyscaling}
\begin{figure}
    \centering
    \includegraphics[width=1.0\linewidth]{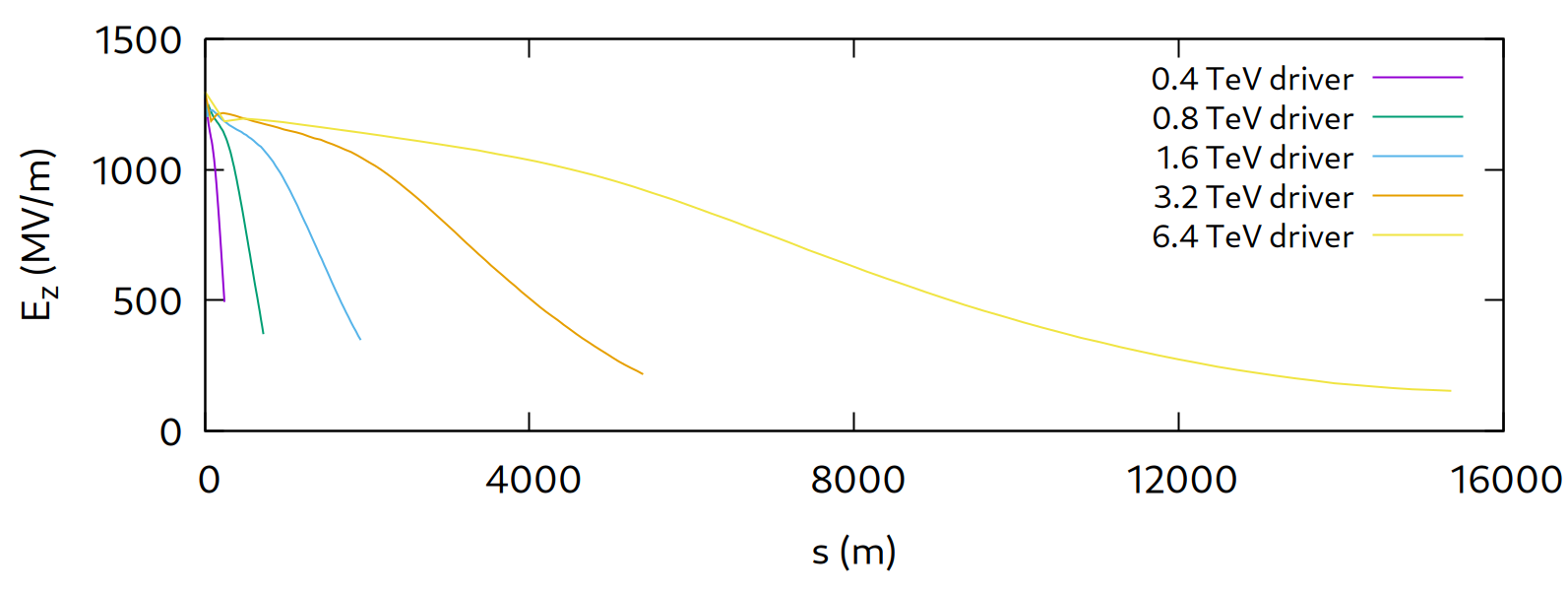}
    \caption{ Evolution of the wakefields as a function of distance for different driver energies.  The distance over which the wakefields maintain their high amplitude increases better-than-linearly with driver energy.}
    \label{fig:EnergyScaling}
\end{figure}

The proof-of-concept simulations carried out for the Higgs factory~\cite{ref:pdwfa-higgs}, discussed in Section~\ref{sec:HiggsPaper}, showed that acceleration was predominantly limited by longitudinal dispersion of the drive bunch.  Increasing the driver energy, $E_d$, reduces dispersion, allowing the wakefield amplitude to be maintained over longer distances, and higher witness energies, $E_w$ to be achieved.  Furthermore, since dispersion scales nonlinearly with energy, the witness energy scales as $E_w\sim E_d^{3/2}$.  Simulations were carried out to investigate the evolution of the unloaded wakefields for different driver energy, with the results shown in Fig.~\ref{fig:EnergyScaling}.  For driver energies up to 1~\si{\tera\electronvolt}, the main limiting mechanism is dispersion, and simulations show excellent agreement with the predicted scaling.  For higher energies, erosion of the head of the drive beam becomes more significant, but this could be overcome through the use of external focussing.  As discussed above, shaping the drive beam would also extend the distance over which the fields maintain their high amplitude.  This better-than-linear scaling with driver energy makes proton-driven acceleration especially attractive for high-energy physics applications beyond the Higgs factory.

\subsection{Energy transfer efficiency}\label{sec:efficiency}
The efficiency with which energy can be transferred from the driver to the witness bunch can be considered in terms of the efficiency with which energy is transferred to and from the plasma.

As the drive bunch propagates through the plasma, it excites a wakefield which is in turn used to accelerate the witness bunch.  The same wakefields act on the driver, causing it to lose energy.  This energy loss ultimately limits the acceleration length, as longitudinal dispersion causes the drive beam to deform.  It is therefore convenient to consider the transformer ratio, $T_R$, given as the ratio of the maximum accelerating field behind the driver to the decelerating field acting on the driver.  For linear wakefields, the maximum transformer ratio is $T_R=2$ for a symmetric driver, and can be increased by using a driver profile which is structured to give a more uniform decelerating field~\cite{pwfa-bane-ramped}, allowing more energy to be extracted from the drive bunch.
%while for a structured driver the transformer ratio is limited by the bunch length~[zholents].  

The energy transfer from the plasma to the witness is limited by the requirement to maintain high accelerating gradients.  100\% energy efficiency would correspond to a witness bunch which fully suppresses the wakefields excited by the driver, in which case the accelerating field acting on the tail of the witness bunch is asymptotic to zero.  This in turn results in either negligible energy gain or the development of a large energy spread.  In order to maintain a large accelerating gradient over the entire witness, the beamloading, i.e.\ the extent to which the wakefields are suppressed, should be limited.  Assuming a witness bunch with a longitudinal profile tailored to ensure uniform acceleration and a low energy spread~\cite{meer-beamloading}, beamloading naturally results in a trade-off between efficiency and gradient.

Beyond these general constraints, the charge and transverse size of the driver and witness also play a role.  For linear and quasi-linear plasma response, the transverse extent of the wakefields depend on the transverse extent of the bunch.  Efficiency is therefore greatest when the driver and witness have the same transverse size.  This limitation is removed in the case where both the driver and witness particles are negatively charged. The drive bunch generates a bubble, and the absence of screening from plasma electrons results in witness wakefields which extend across the width of the bubble, essentially independent of the witness width, making the bubble regime an efficient energy converter from driver to witness.

The nonlinear plasma response to the driver also modifies the transformer ratio.  For a negativity charged driver, the driver charge decreases the local plasma frequency, which increases the transformer ratio.  For a positively charged driver, the opposite is true.%The nonlinear plasma response leads to a slight increase in the transformer ratio for a negatively charged driver, and a slight decrease for a positively charged driver.  

A proton driver therefore has several drawbacks which impact on the energy transfer efficiency.  Protons have a large rest mass, and so longitudinal dispersion of the beam becomes significant at much higher energies than for a lepton driver, limiting the fraction of energy which can be extracted from the drive bunch.  This large mass also leads to a driver with a much larger geometric emittance than the witness, leading to a marked difference in their respective transverse sizes.  Although a positively charged bunch can be used to drive a blowout, the reduction in transformer ratio with increasing nonlinearity makes this scheme unattractive.  The difference in the transverse extent of driver and witness wakefields will therefore act to limit the efficiency.  However, the intrinsic benefits of acceleration in a single stage may outweigh these limitations.

%Despite these limitations, a proton driver still offers several advantages.  Their large rest mass allows protons to be accelerated to high energy in circular machines, and this high energy allows wakefield acceleration to the energy frontier in a single stage.  Avoiding staging not only increases the average accelerating gradient, but also avoids potentially significant energy loss to synchrotron radiation as the witness beam traverses the inter-stage optics.

Proof-of-principle simulations for a Higgs factory~\cite{ref:pdwfa-higgs} (discussed in Section~\ref{sec:HiggsPaper}) using drive and witness bunches with longitudinally symmetric profiles demonstrated an energy transfer efficiency of $\sim3$\%.  By appropriately shaping the driver and witness, this can readily be improved, with shaping of the drive bunch expected to be especially effective due to the impact of longitudinal dispersion.  A detailed optimisation of this scheme is planned, and transfer efficiencies of $\sim10$\% are not unreasonable.  As discussed in Section~\ref{sec:energyscaling}, above, the impact of dispersion is smaller for higher driver energies, and so higher efficiencies can be achieved.  For the case of a 6.8~\si{\tera\electronvolt} driver, discussed in Section~\ref{sec:tenTeV}, the same unoptimised bunch profiles resulted in an energy efficiency of 8\%.  For this case, the efficiency can be increased not only by shaping the bunches, but through the use of external focussing, as erosion of the driver head becomes significant over the longer acceleration length.

\subsection{Plasma}
The generation of hundreds of meters (for a Higgs Factory) or even kilometers (e.g., for a 10~TeV Collider) of seamless plasma with adequate density, uniformity and reproducibility imposes a length-scalable plasma source such as those under development for the AWAKE experiment \cite{ref_Buttenschon_helicon_2018, ref_Torrado_DPS_2023}. 
The direct current Discharge Plasma Source (DPS) \cite{ref_Torrado_DPS_2023} was recently used in the AWAKE experiment to produce plasmas with length up to 10 m that were able to self-modulate a 400 GeV proton beam \cite{ref_turner_ion_motion_PRL_2025_arXiv}. 
The DPS produces the plasmas inside a glass tube filled with a low pressure ($\sim$ 1-20 Pa) gas with an electric discharge between two cold electrodes made of refractory metal. 
A cathode is typically a set two symmetric  metal pins that slightly penetrate into the tube. The current on each pin is forced to be equal by a magnetic choke. The anode can be made of similar pins or of a ring to allow beam propagation on the tube axis. 
The tube has a low capacitance shield  consisting of a symmetric set of wires connected to the anode and  extending close to the cathode. 
The discharge pulses are short (typically a few microsecond) to improve efficiency and avoid plasma self-modulation that would impact on  density uniformity. 
A two-step operation is used to further reduce power. First, the plasma is ignited applying a negative high-voltage (typically 25-50 kV) pulse to the cathode. 
The electric field between the cathode and the shield allows the fast development of a ($\sim$ 10-50 A) arc from the cathode that charges the tube walls, guiding the discharge towards the anode creating a uniform resistive plasma with a low ionisation fraction (typically $<$ 10\%).
The second step consists of a capacitive high current and lower voltage (typically $<$ 10~\si{\kilo\volt}) that heats the plasma until it reaches the desired electron density. The DPS can produce long single plasmas in noble gases with electron densities in the range ($1-50\times10^{14}$ \si{\per\cubic\centi\meter}) with lengths up to 20~\si{\metre} at a reduced cost and using approximately 100 J of electric power per plasma.

The plasma length can be doubled by using a common cathode in the middle of the tube and forcing symmetric tube currents with additional magnetic chokes. Unlimited length scalability can potentially be reached by connecting the double plasma modules through common anodes (this step has not yet been demonstrated).

Applying the DPS to a HEP Collider brings several new challenges that require further R\&D. 
The repetition rate is expected to increase about 5 orders of magnitude from the current repetition rate to the planned 10-20 kHz. 
The bunch being accelerated (probe beam) is expected to become intense and able to produce field ionisation of both neutral gas and plasma ions. This requires a plasma close to fully ionised and the use of a gas with a sufficiently high energy gap between the first and second ionisation levels.
The intensity of the probe beams may produce localised electric fields able to induce fast motion of light background ions that may affect the wakefield structure, this may limit the use of light ions.
The high duty cycle of the DPS is expected to introduce a non-negligible ion drift that may result in uncontrollable axial density gradients. 

\subsubsection{Discharge source with alkali vapor}

Scaling the current Argon/Xenon based DPS to a 1 km plasma operating at 15 kHz would require an electric power on the order of 150 MW. 
Reaching close to full first ionisation will likely require 4 times higher power. The substitution of noble gases by alkali vapors significantly increase the complexity of the DPS but presents two potential advantages: a significant reduction (estimated $\sim$ 10 x) of the electric power to produce similar plasma densities and an increase in the ionisation potential for the second electron which allows the use of more intense beams. 
In such long plasma tubes operating at high duty cycle, further energy saving should be considered. The main energy loss mechanism is the loss of electron-ion pairs to the tube wall. This loss can be reduced and controlled by a solenoid magnetic field able to make the Larmor radius of the ions a small fraction of the tube/plasma radius (typically $0.5$ to $1.5$~T for $> 25$~mm diameter tubes).

We expect the DPS vapour source to operate at a pressure close to 10 Pa and therefore at temperatures close to (612$^{\circ} C$, 344$^{\circ} C$, 265$^{\circ} C$,212$^{\circ} C$, 195$^{\circ} C$) for respectively (lithium, sodium, potassium, rubidium and caesium).
The design of a DPS to operate at these temperatures using materials compatible with the alkaline vapours is further complicated by the presence of the solenoid coils. For the required frequency of operation the solenoid field must be continuous and therefore produced by a superconducting set of coils.

One of the main challenges will be to keep the uniformity of the solenoid magnetic field along the plasma tube specially in the zones were the coils need to be interrupted to source the discharge currents and to renovate the plasma gases/vapors and diagnose the plasma density and particle beam position. 

To reduce the parasitic plasma tube inductance to allow the use of microsecond short pulses, we expect that each plasma section to not be longer than 10 m. This will make a double plasma module up to 20 m long. Therefore, 1 km of plasma will be made of about 50 double plasma modules using 50 bipolar pulse generators and will use 100 main superconducting solenoid coils.

The ion drift can be mitigated or even used to improve the plasma uniformity near the electrodes by employing a bipolar topology for the discharge. For the expected operating frequencies, the plasma only needs an ignition in the first pulse of a burst (since the plasma has no time to fully recombine). Therefore the re-heating pulse can have, with little extra complexity and cost,  two sub-pulses with opposite polarities that result in net zero ion drift.  

\subsubsection{Ion motion} 

Ion motion may be a limiting factor in most plasma acceleration schemes. The electric fields present in the wakefield may be enough to induce fast enough motion in the plasma ions to affect their density and therefore reduce the amplitude of the acceleration/focusing wakefield.
In AWAKE, the long duration of the driving proton bunches imposes the use of plasmas with mass equal or above Argon \cite{ref_vieira_ion_motion_PRL_2012} and recent experiments confirmed that helium ions were removed from the plasma axis by the driving proton bunch suppressing the wakefield while Argon and Xenon were able to keep the wakefield amplitude along the whole driving pulse \cite{ref_turner_ion_motion_PRL_2025_arXiv}. The present scheme uses short driving proton bunches and it is expected that ion motion will be caused by the probe electron bunch particularly if its self-field is enhanced by self-focusing and self-steepening along the acceleration process. This effect may limit the probe beam charge that can be accelerated but can be mitigated by increasing the ion mass. However, increasing the ion mass reduces the gap between the first and second ionisation potentials (e.g. 29.8 eV for Helium and 9.1 eV for Xenon, or 70.2 eV for Lithium and 19.3 eV for Caesium). These two opposing effects (ion motion and field ionisation of the second electron, both due to probe electron beam intensity) are the main factors on choosing the optimum plasma material. 

% end of new plasma section ... NL 

\section{Higgs Factory Study}\label{sec:HiggsPaper}
%In order to demonstrate the utility of proton-driven plasma wakefields, 
A preliminary study was carried out to investigate the potential of our scheme as the basis of a future collider facility~\cite{ref:pdwfa-higgs}. A Higgs factory was chosen, having been identified by the 2020 update of the European Strategy for Particle Physics as the highest-priority next collider~\cite{espp2022}, making the study timely and relevant for the particle physics community.

To effectively excite a wakefield, the drive beam should have a duration on the order of the plasma skin depth, $1/k_p=c/\omega_p$.  A plasma density of \si{\num{3.14e14}~\centi\metre^{-3}} was chosen, corresponding to a skin depth of \si{\num{300}~\micro\metre}, and a driver with a population of \num{1e11} protons, a parabolic profile, and a root-mean-square (rms) length of \si{\num{150}~\micro\metre}.  Longer beams can be facilitated by choosing a lower plasma density, albeit with a commensurate decrease in accelerating gradient.

The driver energy was chosen as 400~\si{\giga\electronvolt}, corresponding to the proton beam from the CERN Super Proton Synchrotron (SPS) used in AWAKE.\footnote{Note that a 500~\si{\giga\electronvolt} driver was assumed in the development of the proton acceleration facility discussed in Section~\ref{sec:protons}.}  An initial rms energy spread of 10\% was assumed, in order to keep the longitudinal emittance in line with currently achievable beam parameters.  As the protons in the drive bunch are much more massive than the leptons in the witness, the relative velocity of the two becomes significant over the acceleration distances required to reach high energy.  In order to prevent dephasing, a plasma density gradient is employed, as discussed above in Section~\ref{sec:shortdriver}, which keeps the relative phase of the driver and witness constant as the absolute distance between the two decreases.  The witness bunch is injected into the second period of the wakefields in order to reduce the required variation in density.  A schematic of the initial conditions as the driver and the electron witness enter the plasma are shown in Fig.~\ref{fig:PDPWFA_clipart}.

In addition to the strong accelerating fields which plasma provides, beams in plasma are also subject to strong focussing fields.  In order to prevent the rapid pinching of the drive beam, a tailored emittance profile was used, with the emittance increasing monotonically along the length of the beam. Further studies are planned on how such a profile can be achieved, with the exploitation of plasma instabilities (similar to the AWAKE scheme) a promising candidate.  A correctly prepared beam was shown to allow stable propagation in plasma over the distances required for acceleration.

Simulations of the acceleration process demonstrated that a 400~\si{\giga\electronvolt} drive beam was sufficient to accelerate witness beams to the energies required for a Higgs factory, with bunches of \num{1e10} electrons and positrons accelerated to 125~\si{\giga\electronvolt} in 200~\si{\metre} of plasma.  The main limitation on the acceleration length is due to longitudinal dispersion of the drive beam, arising due to the correlated energy spread which the drive bunch develops as it loses energy to the plasma, with very little contribution from the initial uncorrelated spread.  As the drive bunch lengthens, it no longer effectively drives a wakefield, reducing the accelerating gradient.  Shaping of the drive bunch is a known method to improve the transformer ratio of wakefield accelerators~\cite{meer-beamloading}, and should be further investigated in the context of proton-driven schemes.

The witness bunches were also chosen to have an initially parabolic longitudinal profile, resulting in an accelerated electron bunch with the ``quasi-monoenergetic'' spectrum which is characteristic of wakefield accelerators.  This arises due to the plasma response to the witness bunch itself, which modifies the accelerating field which acts upon it.  This can also be controlled by shaping the bunch profile, and has been shown to be an effective method to control the final energy spread~\cite{lindstrom-energyspread}.  Controlling beamloading in this way also allows energy to be extracted from the wakefields more efficiently, and so is also expected to increase the charge which can be accelerated.  In the absence of ion motion, the electron emittance is preserved during acceleration.  Initial studies indicate that when ion motion is included, the emittancence growth remains rather small due to the slow evolution of the focussing fields over the acceleration length~\cite{pwfa-farmer-ion_selfmatching}.

The acceleration of positrons remains challenging, with simulations showing that the required energy can be reached, but that the beam quality is significantly impacted during acceleration.  As noted in Section~\ref{sec:positrons}, there is an ongoing effort across the plasma wakefield community to develop new techniques for positron acceleration.  We also note that some schemes for future colliders would avoid the need for plasma acceleration of positrons, such as $\gamma\text{--}\gamma$ or $e^-\,e^-$~\cite{IP-zhang-anomalous_electron_pinch} colliders, or asymmetric schemes~\cite{ref:HALHF,EPPSU2026_HALHF}.

The study made several assumptions for the achievable parameters, notably on the proton drive beam, the plasma source, and the acceleration of positrons.  As further studies on these aspects of the scheme are carried out, the estimates can be further refined.  Table~\ref{tab:Table-1} represents the most up-to-date values, with the main assumptions discussed in Section~\ref{sec:IP}.

% \subsection{Injector for FFC? \emph{JF}}
% Maybe remove this, see the section on injector for EIC

\section{Applications}

We discuss further applications beyond the initial Higgs Factory study, starting with the need for a demonstrator project.

\subsection{Development of a demonstrator}
Given the novelty of proton-driven wakefield acceleration, a demonstrator facility should be developed before embarking on the construction of collider facilities.  The realisation of such a facility would provide an experimental proof-of-principle, and allow key issues such as reliability to be investigated, but crucially must have real physics applications.

An electron booster for the EIC injector has been identified as a potential candidate, as the EIC facility already has access to proton bunches at 250~\si{\giga\electronvolt}.  The envisaged booster would accelerate 400~MeV electrons to 18~\si{\giga\electronvolt}.  Although significant development of the accelerator ring would be required to allow the acceleration of short proton bunches, the scheme could be competitive with a conventional RF booster.  A detailed analysis based on the ideas detailed above is currently underway.  A previous study based on the AWAKE scheme was reported in~\cite{Chappell:2019ovd}.

\subsection{Electron-proton and electron-ion collider}
An electron acceleration scheme based on PDPWA would in principle allow for much higher energy eP and eA collisions than previous or future ring-based schemes, and would be an ideal %first implementation
intermediate scheme for PDPWA as positrons are not required and much reduced specifications are possible on the electron bunch.  The accelerated electrons could be collided with, e.g., LHC protons and ions to produce very high energy eP and eA collisions.  As an example, we assume that the SPS and its pre-injectors have been retrofitted with the Fixed-Field Alternating Gradient (FFAG) Accelerator
scheme  described above.  Such an FFAG scheme is expected to yield short proton bunches of $10^{11}$ protons/bunch and up to $E_p=500$~\si{\giga\electronvolt} per proton at a rate up to 20~kHz.  With protons of this energy, we expect that  bunches of $2\cdot 10^{10}$ electrons or positrons can be accelerated up to $E_e=200$~\si{\giga\electronvolt} in a plasma section %\textcolor{red}{!!!simulations don't support this - the scaling laws suggest a 600~\si{\giga\electronvolt} driver would allow acceleration to 230~\si{\giga\electronvolt}}.  
yielding $E_{\rm cm}\simeq 2360$~GeV for collisions with LHC protons. The emittance of the electron bunches will be adequate such as to allow a match of the transverse dimensions to the proton bunch size at the IP.  Using the proton bunch parameters from the LHeC studies, a luminosity of $10^{32}$~cm$^{-2}$s$^{-1}$ will be possible. The kinematic range that will become available for DIS experiments using the SPS as proton drive beam will be extended by approximately two orders of magnitude relative to HERA.  If eventually also the LHC tunnel can be used to house the proton driver accelerator, considerably higher final lepton energies will be available, and CoM energies of $E_{\rm cm}=9000$~GeV can be reached~\cite{Caldwell:2016cmw}.

\subsection{High-energy \texorpdfstring{$e^+e^-$}{e+e-} Colliders}\label{sec:tenTeV}

Proton bunches of $550$~\si{\giga\electronvolt} would allow for the acceleration of electron bunches in plasma to $275$~\si{\giga\electronvolt}, and such proton bunch energies could be achieved in the SPS, Tevatron, HERA or even possibly the RHIC/EIC tunnels. 

Looking to higher energies, the recent Particle Physics Project Prioritization Panel in the US~\cite{P5} called for an extensive R\&D programme that would allow the development of a 10~\si{\tera\electronvolt} parton centre-of-momentum collider.  Preliminary simulations show that a 6.8~\si{\tera\electronvolt} driver, equivalent to the energy of the LHC, with \num{1e11} protons can accelerate \num{1e10} electrons to 5~\si{\tera\electronvolt}.  Additional studies currently underway, including bunch shaping and external focussing, offer a path to further increase the efficiency.

The inherent trade-offs introduced by beamloading, discussed in Section~\ref{sec:efficiency}, mean that such a collider would also offer the option of higher-charge witness bunches accelerated to lower energy.  This would allow the physics at lower energies, for example the range up to 1~\si{\tera\electronvolt}, relevant for Higgs couplings~\cite{EPPSU2026_LCVision}, to be explored at much higher luminosity.

% \begin{figure}
%     \centering
%     \includegraphics[width=0.5\linewidth]{10TeV.png}
%     \caption{Enter Caption}
%     \label{fig:10TeV}
% \end{figure}

\section{Conclusion}

We have discussed the main elements of a proton-driven plasma wakefield accelerator based collider facility.  To achieve interesting luminosity, the proton driver will need to deliver short ($\ll 1$~mm) proton bunches at high rates.  We reviewed two schemes for achieving this - a rapid cycling synchrotron and an FFAG-based scheme, and conclude that the FFAG scheme will provide higher luminosity.  A first self-consistent set of parameters was developed indicating that such a scheme is indeed potentially feasible.  There are clearly many challenges to the development of this scheme, including novel RF acceleration modules and high precision and strong magnets.

We further discussed the plasma acceleration section, indicating where the main challenges lie.  These are identified as 1) the ability to accelerate positrons while maintaining necessary emittance and 2) improving the energy transfer efficiency.  Challenges also appear in the development of appropriate plasma cells, as we report here.

Many exciting applications would become available should a scheme such as ours become a reality. Given the potential for such accelerator facilities, these challenges should be taken up and met.

\section{Acknowledgments}
A.P. acknowledges support from BMBF-Projekt 05P24PF2 in the preparation of this document.
\newpage

\onecolumngrid
\makeatletter
\renewcommand{\@title}{}
\renewcommand{\@date}{}
\@author@finish
\titleblock@produce
\makeatother
\twocolumngrid
% \section{References}
\bibliography{Higgs}
% \printbibliography[heading=none]

\end{document}